# Using Shape Diversity on the way to new Structure-Function Designs for Magnetic Micropropellers

*Felix Bachmann*[1]*, Klaas Bente*[1,2]*, Agnese Codutti*[1,3]*, Damien Faivre*[1,4]*

[1] Department of Biomaterials, Max Planck Institute of Colloids and Interfaces, Science Park Golm, 14424 Potsdam, Germany
[2] Department of Nondestructive Testing, Bundesanstalt für Materialforschung und -prüfung, Unter den Eichen 87, 12205 Berlin, Germany
[3] Department of Theory & Bio-Systems, Max Planck Institute of Colloids and Interfaces, Science Park Golm, 14424 Potsdam, Germany
[4] CEA/CNRS/ Aix-Marseille Université, UMR7265 Institut de biosciences et biotechnologies, Laboratoire de Bioénergétique Cellulaire, 13108 Saint Paul lez Durance, France

## Abstract

Synthetic microswimmers mimicking biological movements at the microscale have been developed in recent years. Actuating helical magnetic materials with a homogeneous rotating magnetic field is one of the most widespread techniques for propulsion at the microscale, partly because the actuation strategy revolves around a simple linear relationship between the actuating field frequency and the propeller velocity. However, the full control of the swimmers' motion has remained a challenge. Increasing the controllability of micropropellers is crucial to achieve complex actuation schemes that in turn are directly relevant for numerous applications. The simplicity of the linear relationship though limits the possibilities and flexibilities of swarm control. Using a pool of randomly-shaped magnetic microswimmers, we show that the complexity of shape can advantageously be translated into enhanced control. In particular, directional reversal of sorted micropropellers is controlled by the frequency of the actuating field.

This directionality change is linked to the balance between magnetic and hydrodynamic forces. We further show an example how this behavior can experimentally lead to simple and effective sorting of individual swimmers from a group. The ability of these propellers to reverse swimming direction solely by frequency increases the control possibilities and is an example for propeller designs, where the complexity needed for many applications is embedded directly in the propeller geometry rather than external factors such as actuation sequences.

## I. Introduction

Microswimmers are envisioned for a multitude of applications ranging from solving environmental problems to being used for micro surgery [1-3]. Precise, versatile and non-invasive controllability is necessary to cover this broad scope of applications. These requirements are mostly matched by magnetic microswimmers. The fuel-free actuation by weak and homogeneous magnetic fields indeed allows remote controlling in many environments, the synthesis via nanofabrication makes them accessible even on a sub-micrometer scale [4-6]. In addition, the ability to functionalize their surface and the limited toxicity of the mostly iron-based propellers makes them appealing for medical applications [2,7]. Many of the current magnetic microswimmers use a helical shape with a fixed magnetic moment to rotate in an externally applied magnetic field, which enables stable propulsion. In this case, a simple linear relationship between the frequency of the actuating magnetic field and the velocity of micropropellers is used to precisely control the propeller [5,8-10]. This leaves the sign of the swimming direction of the propeller to be determined by the rotation direction of the applied magnetic field, which limits the versatility of their actuation capability: when controlling two or more geometrically identical propellers, it is not possible to let them swim in a common propulsion mode respectively in the same direction and, if needed, in opposite directions, simply because they identically react to the same rotation and / or rotation reversal of the field such that they eventually all swim always in the same direction. This does not change even by using the non-linear propeller behavior after the so called stepout frequency.

Joining and splitting of swarms of microswimmers in 3D plays however an important role for multi targeting tasks from micro-manipulation to self-organization and drug delivery. With the helical swimmers, this is only possible with complex actuation sequences [11] that have remained theoretical. Recent studies have now shown that non-linear propelling behaviors can be obtained for particular devices [12-15]. The general theory describing linear and non-linear cases depicts a change of the propeller's axis of rotation as a function of the externally applied frequency for many geometries[16]. So far, this behavior was only appreciated as a non-swimming (tumbling) and a swimming (wobbling) regime [12,14].

Here, we take advantage of a synthetic route to random-shaped micropropellers [7] to test new actuation schemes in this context. Screening a pool of randomly shaped micropropellers, we select those reversing their swimming direction based on the applied actuation frequency. Comparing this frequency-induced reversal with recent progresses in their theoretical description, we make an argument for expanding the degree of controllability of micropropellers. We in particular demonstrate the isolation of a single propeller from a swarm. This structure function relationship can lead to a new direction in designing magnetic micropropellers, where controllability is not embedded into actuation sequences but is already included in the geometry of the microswimmer.

## II. Results

### A. Direction reversal and correlation with propeller orientation

Randomly shaped magnetic micropropellers were observed in an inverted custom-designed optical microscope [7,17] and are actuated in water far away from surfaces by a rotating magnetic field $B(\omega,t) = [B_0 \sin(\omega,t), 0, B_0 \cos(\omega,t)]'$. Some of them show a behavior we call frequency-induced reversal of swimming direction (FIRSD). In other words, the propeller swims in two opposing directions for two different field frequencies of the external actuating magnetic field, while no other parameter is changed. This stands in contrast to direction inversions used before, where the magnetic field rotation needed to be inverted to reverse the swimming direction. The velocity-frequency-relationship of propellers exhibiting a FIRSD is presented in **Figure 1 (A and C)** for two exemplary propellers. An inverse radon-transformation provides 3D-reconstructions of the propeller shapes from the recorded 2D images using the projections of the rotating propellers (**Figure 1 B and D**, more information in SM, Fig. S6 and S7) [18,19]. Additionally, microscope image snapshots of the propeller configuration to the horizontal axis of rotation are put in the according frequency regime at the bottom. A theoretical fit (see SM section I. and II.) is added to the experimental results of the frequency-dependent velocity measurements in Figure 1 A and C. Typically, the curves exhibit a linear regime at low frequencies where the propellers rotate around their short axis (tumbling, region I in Figure 1) (propeller 1: f = 0 - 25 Hz; propeller 2: f = 0 - 68 Hz). At a transition frequency $f_{tw}$, the behavior changes, and the propellers tilt their long axis towards the horizontal axis of rotation of the actuating field (wobbling, cf. **Figure 2** A) (propeller 1: ≈ 25 Hz; propeller 2: ≈ 68 Hz). In this frequency regime (region II in Figure 1), the fit as well as the experimental data imply the existence of two solution branches (blue solid line and green dashed line). These correspond to different orientations of the propeller during its actuation in an external rotating field. First, we focus on one branch (solid blue) and will discuss the implications of the branching in the next section. The change in the axis of rotation with respect to the propeller geometries alters the rotation-translation coupling of the propellers at each frequency step. In general, this leads to a non-linear relationship between the actuating field frequency and the propeller velocity. Here, the sign of the coupling and therefore the velocity changes with frequency: both propellers slow down after $f_{tw}$ and eventually reverse their swimming direction (propeller 1: ≈ 27 Hz; propeller 2: ≈ 77 Hz). Due to the different orientation of the propeller towards the axis of rotation, the behavior deviates from the single linear behavior seen for other magnetic micropropellers [5,9,10,20] : while still turning in the same sense of rotation, the swimming direction reversed. Finally, the velocity breaks down at the frequency $f_{so}$ (propeller 1: ≈ 55 Hz; propeller 2: ≈ 117 Hz), reminiscent to what is seen for the behavior after the step out frequency of the linear propellers. Here, the propellers, with their magnetic moment *m* fixed in their geometry cannot follow the frequency of the magnetic field rotation anymore and apparently return to a rotation around their short axis (microscope images panel A and C, asynchronous regime, region III in Figure 1). The maximum velocities (Table I) for propeller 1 for the two opposing swimming directions are $v$ = 3 µm s$^{-1}$ at $f$ = 22 Hz and $v$ = -9 µm s$^{-1}$ at $f$ = 41 Hz. Propeller 2 swims at $v$ = -11 µm s$^{-1}$ for $f$ = 66 Hz and up to 24 µm s$^{-1}$ for $f$ = 117 Hz. The respective minimal and maximal dimensionless velocity $U = 1000 \cdot v/(l \cdot f)$ of the two propellers are also shown in Table I, where $f$ is the frequency of the actuating external field and $l$ the characteristic length of a propeller. The lowest and highest dimensionless speeds for propeller 1 are about 36 and -52 respectively, for propeller 2, they are 51 and -41. In summary, both propellers reorient themselves depending on the frequency and are able to effectively swim in two opposing directions by only changing the applied external frequency of the magnetic field.

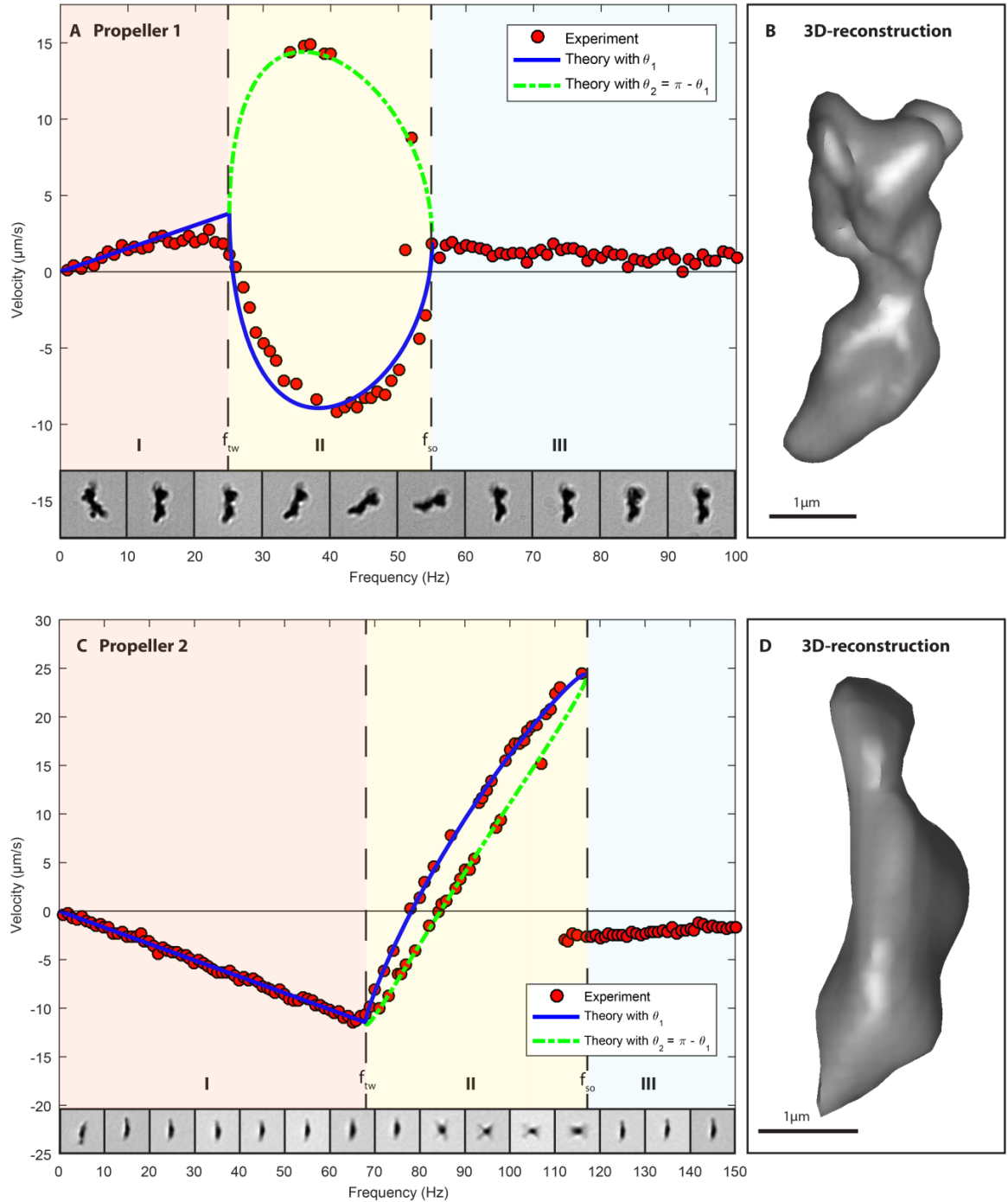

**Figure 1. Velocity-frequency-relationship.**
The measured velocity (red dots) dependent on the applied frequency of the actuating magnetic field is shown in a panel **A** and **C** for propeller 1 and 2 respectively. The theoretic fit on those data points is shown in blue (solid line) and green (dashed line), which also describes the observed branching (corresponding to different propeller configuration at the respective frequencies). The 3D-reconstruction of the propellers is depicted in panel **B** and **D**. To visualize the change in propeller configuration to the horizontal axis of rotation dependent on the frequency, microscope images are shown in the measured frequency regime for one respective branch (Panel **A** and **C** on the bottom). The most left image corresponds to the orientation for a constant field in between the measurements. Both propellers show a reversal of the velocity direction in the first branch (blue), when increasing the frequency: after a linear tumble regime (**I**, swimming in one direction) until $f_{tw}$, a non-linear wobbling regime (**II** happens up to the step-out frequency $f_{so}$ (**III**). During the wobbling, the long axis of the propeller starts to tilt more and more towards the horizontal axis of rotation and at a certain frequency, the propeller velocity reverses its sign – the propeller swims in the opposing direction.

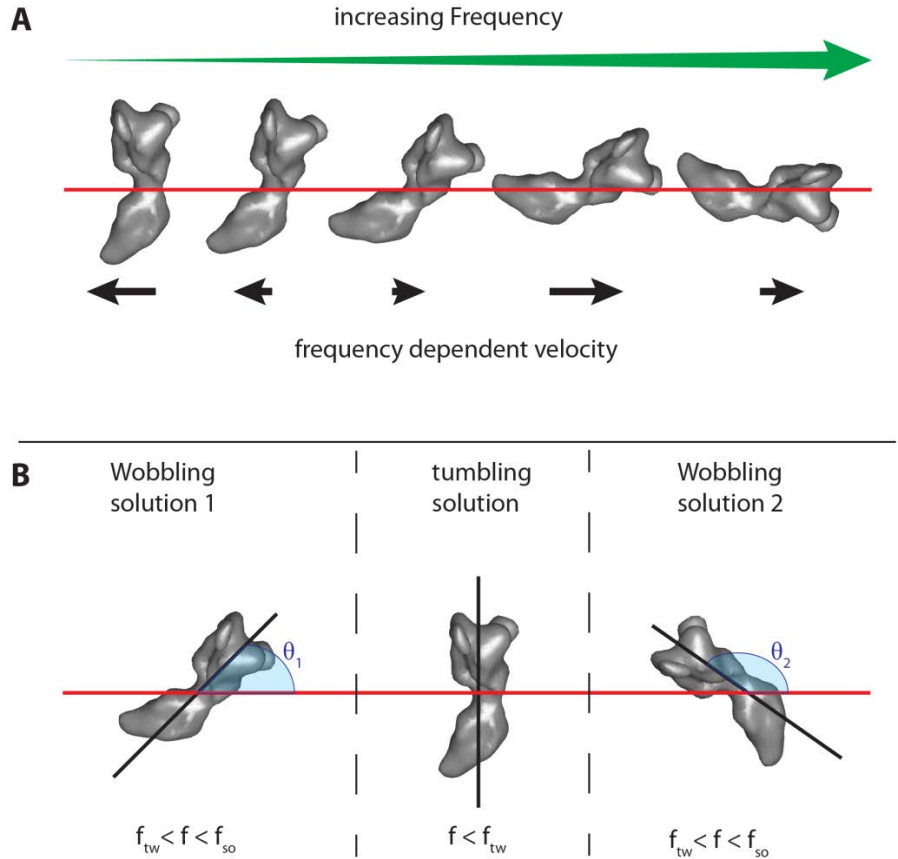

**Figure 2. Frequency dependent axis of rotation (A).**

The orientation of a propeller towards its axis of rotation (red horizontal line) changes with increasing frequency for some geometries to balance the magnetic and hydrodynamic torques. As a result the coupling and therefore the velocity changes with frequency, which can lead to an inversion of the propulsion direction. **Branching schematic (B).** Coming from the tumbling configuration at low frequencies (middle), two wobbling configurations are possible for higher frequencies between the transition frequency $f_{tw}$ and the step-out frequency $f_{so}$. Here the two possibilities of Propeller 1 are depicted (cf. SM, movie S7 and S8): the first wobbling solution with an angle of $\theta_1$ between the long axis of the propeller and the axis of rotation (red horizontal line) and the second solution/configuration (right side) with $\theta_2 = 180° - \theta_1$.

**Table I. Propeller/Measurement characteristics.**

| Property | Propeller 1 | Propeller 2 |
|---|---|---|
| Magnetic field $B_0$ (mT) | 2 | 1 |
| Propeller length $l$ (µm) | 4.34 | 4.13 |
| Propeller diameter $d$ (µm) | 1.59 | 1.28 |
| Approx. magnetic moment[a] $m$ (A m²) | $1.08 \cdot 10^{-14}$ | $3.84 \cdot 10^{-14}$ |
| Magnetic Saturation[b] $m/m_{sat}$ | 0.003 | 0.02 |
| $f_{tw}$ (Hz) | 25 | 68 |
| $f_{so}$ (Hz) | 55 | 117 |
| $v_{min}$ (µm s$^{-1}$) | -9 | -11 |
| $v_{max}$ (µm s$^{-1}$) | 3 | 24 |
|  | 15[c] |  |
| $U_{min}$ | -52 | -41 |
| $U_{max}$ | 36 | 51 |
|  | 98[c] |  |

[a] Calculation from cylindrical approximation [13,21]; [b] with saturation magnetization of maghemite [22]; [c] Second branch values

## B. Torque balance determines propeller reorientation and branching

The reorientation of the propeller with changing frequency can be explained by a balance of the acting magnetic and hydrodynamic torques [13,14] and was recently expanded to arbitrarily shaped particles [16]. When applied to our system, this leads to the following qualitative physical description: at low frequencies, the magnetic moment of the propeller is in the plane of the rotating magnetic field but lags behind with a constant angle/phase as result of the torque balance. With increasing frequency this phase is increasing too. This results in a linear regime (region I in Fig. 1, tumbling). Depending on the geometry of the propeller and the associated magnetic moment, a certain transition frequency $f_{tw}$ exists, where it is favorable to change the propeller orientation and to not rotate around the (hydrodynamically worse) short propeller axis anymore [14]. At this point, the magnetic moment is moved out of the magnetic field plane (non-linear regime, region II in Fig. 1, wobbling), the hydrodynamic drag is decreased by rotating around an axis with lower hydrodynamic viscous drag and the torque balance is restored. This balance (synchronous regime, region I+II in Fig. 1) can only be maintained up to a certain step-out frequency $f_{so}$, where neither the phase-lag can be increased, nor the hydrodynamic drag can be decreased by reconfiguration – the propeller can no longer follow the magnetic field and the velocity breaks down (asynchronous regime, region III in Fig. 1) [14]. A detailed theoretical description of this process using the rotational mobility matrix **F** and coupling mobility matrix **G** together with the Euler angles $\phi$, $\theta$ and $\psi$ to calculate swimming velocities can be found elsewhere[16] but is additionally shortly summarized in the SM section I. and II.

Looking back at the measured data in Fig. 1 between $f_{tw}$ and $f_{so}$, another mutual but unequally spread feature occurs, with some measuring points not following the general curve (also called branching). This branching, which was theoretically predicted [16], is not necessary for the above described frequency-induced reversal of swimming direction, but can offer further possibilities but also challenges (SM). Branching is explained by the fact that two solutions are possible between the two characteristic frequencies $f_{tw}$ and $f_{so}$ for the according Euler angles when applying a frequency above $f_{tw}$: $\theta_1$ corresponds to the blue solid line in Fig. 1 A and C and $\theta_2 = 180° - \theta_1$ to the green dashed line. These two configurations can be seen in the supporting movie S7. Figure 2 B schematically illustrates the two possible solutions at frequencies above $f_{tw}$ when originating from a rotation around the short propeller axis for frequencies below $f_{tw}$. The two different solutions can result in two rather different velocity responses (propeller 1) dependent on the geometry and the associated magnetic moment [16]. Alternatively, propeller 2 shows a configuration where they are very similar. It is noteworthy that even identical velocity-responses were predicted for the two orientations [16], which would avoid non-unique velocities responses in this frequency regime. This complex behavior can be envisioned to obtain up to three different velocity responses (e.g. negative, positive, zero) for a narrow regime of the applied frequencies by providing the right initial conditions.

**C. Changing propeller direction by varying the applied field strength**

An additional alternative to achieve a reversal of the swimming direction is to change the strength of the applied magnetic field instead of its frequency as shown for propeller 1 at three different field strengths (0.5, 1 and 2 mT, **Figure 3 A**). The magnetic torque determining the propeller configuration scales linearly with magnetic moment of the propeller but also with the applied magnetic field strength ($\tau_m = B \times m$). Therefore the characteristic frequency-velocity curves scale with the applied magnetic field as can be seen in the inset of Figure 3, where the frequencies and the velocities of the three measurements at the different magnetic fields are normalized on the respective magnetic field – the curves fall onto each other. This can be used to reverse the swimming direction by only changing the applied magnetic field strength at a constant frequency as illustrated in the theoretical plot of the propeller velocity as function of the applied magnetic field strength and frequency (**Figure 3 B**). This can be achieved because the different regimes (and therefore different swimming directions) can either be reached by changing the applied frequency (horizontally) or the magnetic field strength (vertically). Exemplarily, the propeller has a velocity $v \approx -5$ µm s$^{-1}$ at an external frequency of 20 Hz and a field of 1 mT. At the same frequency but at 2 mT, the velocity of the same propeller is $v \approx 2.5$ µm s$^{-1}$.

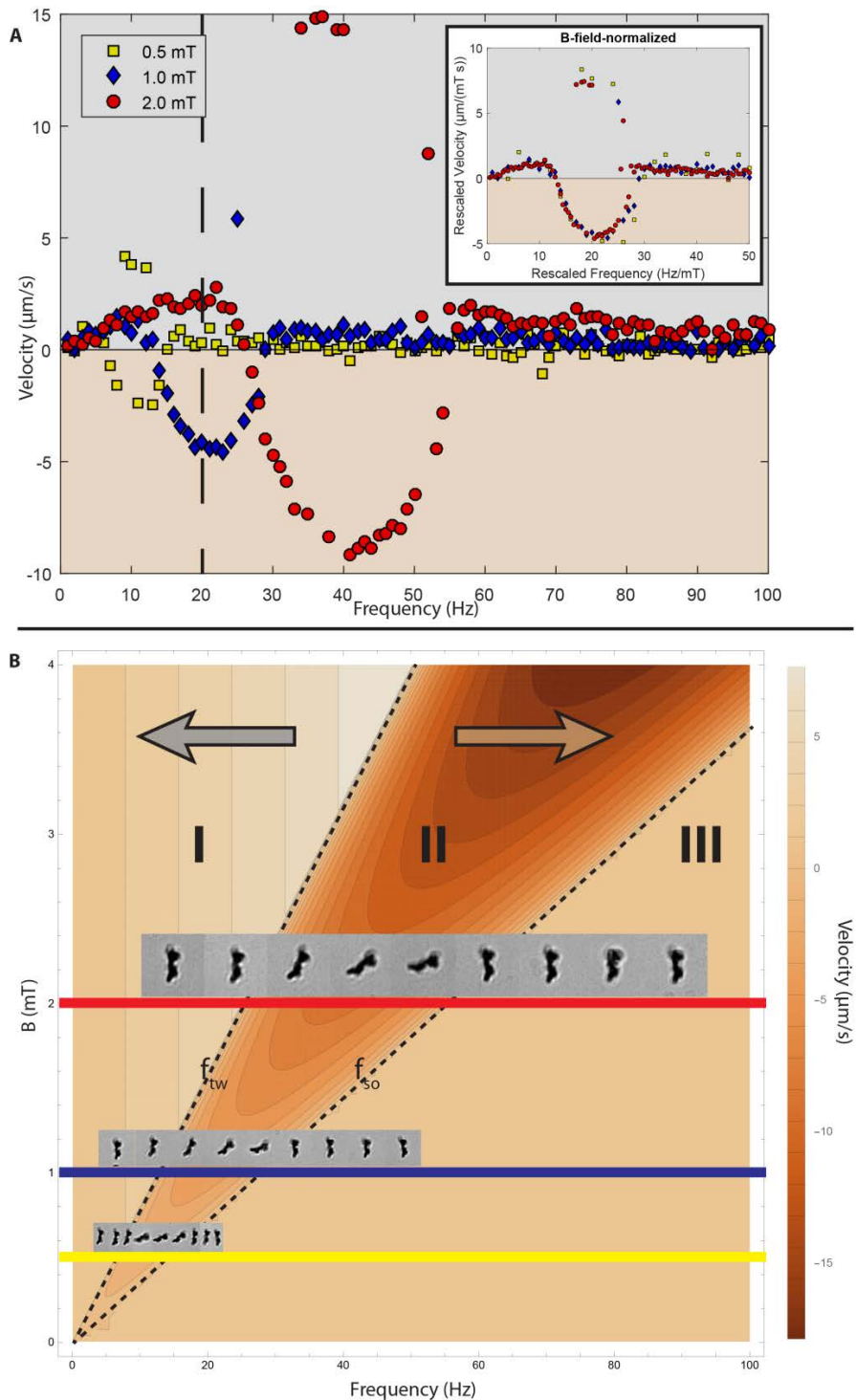

**Figure 3. Scaling with the magnetic field.**
**A**: Velocity-frequency dependence of propeller 1 at three different magnetic field strength (0.5, 1 and 2 mT). Depending on the magnetic field, the propeller can swim in opposing directions for the same frequency, e.g. at f = 20 Hz for 1 mT (blue diamonds) and 2 mT (red dots). This is possible since the frequency characteristics scale with the strength of the magnetic field. In the inset the three curves are normalized on the respective field strength and therefore fall onto each other. **B**: Theoretical velocity as function of the applied magnetic field strength and frequency.

The characteristic frequencies ($f_{tw}$ and $f_{so}$, dashed lines) and therefore the propeller behavior scales linearly with the magnetic field strength. The three measurements at 0.5, 1 and 2 mT are depicted through the yellow, blue and red lines, respectively. Additionally, microscope images of the propeller configuration are shown to illustrate the three different regions and the two arrows indicate the respective swimming direction: the linear regime (I), the wobble regime (II) and the regime after the step-out frequency (III, not included in theory, set to 0).

**D. Implication of FIRSD for swarm control**

The rather basic question is why this behavior offers any benefit when comparing it to a simple reversal of the sense of rotation of the external field that would also lead to a reversal of the swimming direction. The differences between both methods can be best seen, when considering multiple microswimmers actuated by the same external field. Future applications will include tasks like isolating one specific propeller from a swarm of microswimmers. Using previous methods, this is rather complicated but possible by applying a sequence of different actuating field frequencies and directions and using e.g. the step-out behavior after the linear frequency-velocity relationship. However, this currently has mostly remained theoretical or was performed on or close to surfaces in quasi 2D [11,23,24]. A propeller with frequency-induced reversal of swimming direction allows direct targeting of this propeller at a certain frequency. Such a behavior is shown for a swarm of randomly shaped micropropellers in **Figure 4** A and movie S11 in 3D far away from surfaces: at a frequency of 20 Hz, the whole swarm swims in one direction (here left) until the frequency is increased to 40 Hz where all but one propellers continue swimming in the initial direction, the final one reversing its swimming direction. This proof-of-principle motivates a more sophisticated and reliable way, which is schematically depicted in **Figure 4** B: four geometrically identical propellers that show FIRSD only differ in the modulus of their magnetic moment, for example swimmers prepared by 3D printing with varying amount of magnetic materials. While at low frequencies, the propeller with the smallest magnetic moment (and therefore the smallest characteristic frequencies) still swims in the same direction as the rest of the swarm, applying a frequency that is characteristic for this propeller directly reverses its swimming direction, while the rest of the swarm still moves in the original direction (with slightly different velocity). The individual addressed propeller is separated from the rest. This behavior might be expanded to more propellers in the swarm so that for different propeller-specific frequencies, they reverse their swimming behavior compared to the rest of the propellers. This allows joint movement for certain frequencies, agglomeration in a certain point and splitting into groups for other frequencies. Alternatively, this method can be used to collect propellers with a certain desired behavior at selected actuation frequencies from the pool of synthesized randomly particles, sperating them locally.

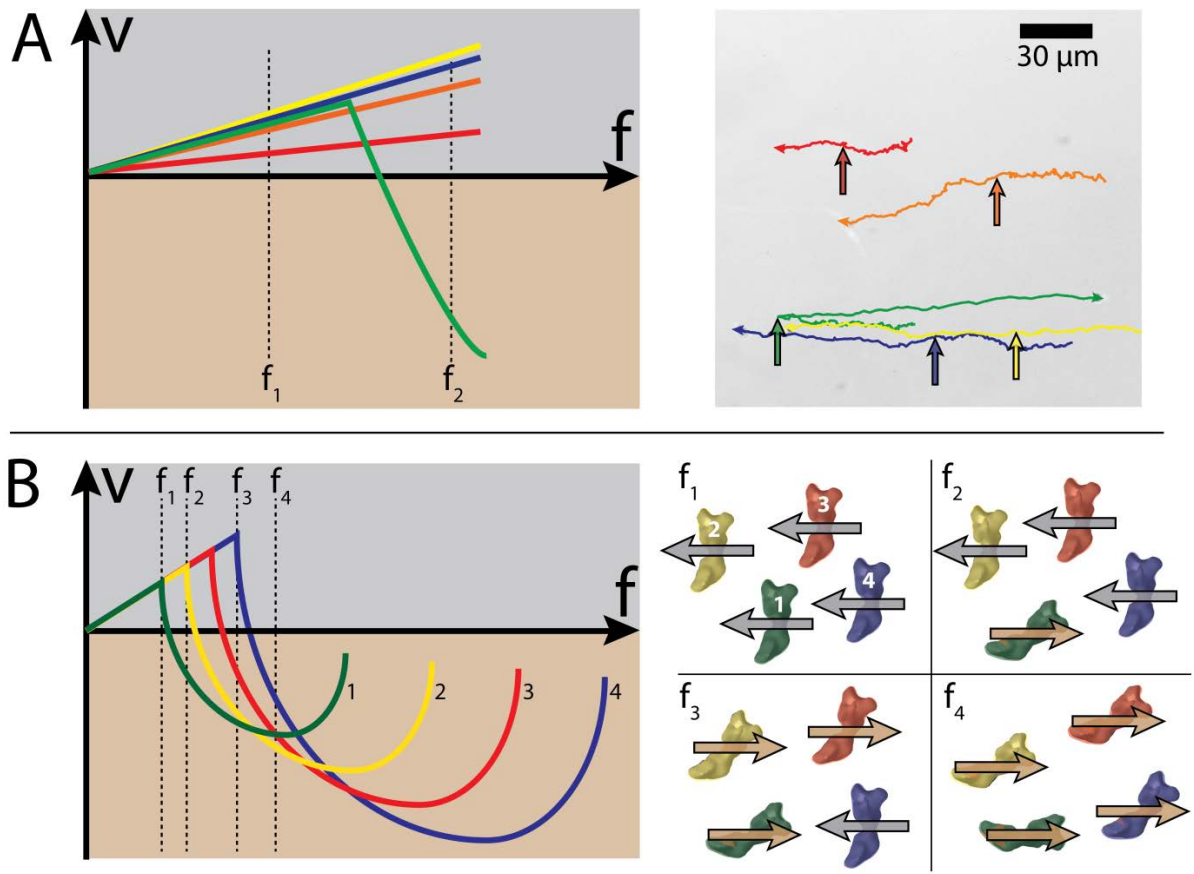

**Figure 4. Propeller isolation for randomly shaped propellers (A).** Tracks of a group of randomly shaped propellers. They swim all to the left for the initially applied magnetic field

frequency of $f_1 = 20$ Hz. The arrows mark the location when the frequency is increased to $f_2 = 40$ Hz. One propeller (green track) shows a reversal of swimming direction at this frequency and swims in the opposing direction, while the rest of the swarm propellers are still in their linear regime and continue to swim to the left. This allows a joint propulsion for one frequency and an isolation of a single propeller for another as can be seen in the supplementary movie S11. The according v-f-diagram is schematically shown on the left. **Propeller selection schematic for identically shaped propellers (B).** Isolation of a single propeller from a swarm of identical propellers that only differ in their magnetic moment modulus (increases from 1 to 4). At a frequency $f_1$ all propellers swim in the same direction. At frequency $f_2$ the propeller with the smallest magnetic moment (green, 1) has already changed its axis of rotation and has a negative velocity, while the rest of the swarm continues swimming in the original direction (positive velocity). At frequency $f_3$ all propellers except for the one with the highest magnetic moment (blue, 4) have reversed their swimming direction. At frequency $f_4$ all propellers swim in the opposing direction compared to $f_1$. This offers new possibilities for controlling swarms of propellers and single propellers at the same time.

## III. Discussion

We showed that it is possible to reverse the swimming direction of magnetic micropropellers by only changing the applied actuating frequency. This behavior offers an alternative to the traditional reversing of the actuating magnetic field and enables simple execution of previously more complicated tasks as it can help shifting the complexity of actuation to the details of the propeller geometry. In swarm control, a group of (non-identical) propellers swims together at a certain frequency. Although the common linear micropropellers are easily reversed with by a rotating field turning in the opposite direction and therefore offer great possibilities to globally manipulate swarms of microswimmers, they need a sequence of at least two magnetic field steps to join or separate subgroups of those propellers [11,24,25]. FIRSD-propellers are usually able to do those basic tasks in one step (SM section IX.). For more complicated applications with longer actuation sequences and more tasks a factor of 2 can rapidly add up. Although FIRSD will not be the optimal solution to all future challenges, they offer new avenues to master them. Recently, there have been other microswimmer systems enabling direction reversal. Gomez-Solano *et al*. reported on thermally induced direction and motility change of self-propelled colloids, mimicking the run-and-reverse behavior of some bacteria [26]. However, the dependency on the surrounding fuel and the activation through laser light makes *in-vivo* applications hard to realize. Khalil *et al.* produced two tailed-microbots that showed frequency dependent back and forth movement through flagellar propulsion [27]. These microbots are still several hundreds of micrometer and might face micromechanical problems when scaling them down, similar to the theoretically proposed equivalent for joint helices [28]. An interesting system was presented by Garci Torres et al. with frequency induced reversal based on an interplay of magnetic, hydrodynamic and gravitational interactions [29]. However, their system is a self-assembled swimmer that requires nearby surfaces to break symmetry and is thereofore rather complemental than alternative to our propellers swimming far away from surfaces. Our proposed system uses penetrable remote control through weak homogenous magnetic fields together with the intrinsic rich interplay between a fixed geometry and magnetism in magnetic micropropellers. This enables a defined change of axis of rotation in dependence of the applied external frequency or magnetic field strength and can therefore be envisioned as an alternative to the common field reversal in some applications. However, there will always be a trade-off between a certain functionality or application a propeller can be used for and its effective propulsion. This is the case for the here reported propellers: more efficient, constant stable propulsion over a large frequency regime is traded with the ability to control the sign of the swimming direction and the propeller speed by applying the right frequency beyond the usual linear relationship. This feature is currently under-used but can add to the controllability of the propellers apart from the currently considered options (field orientation/shape deformation/soft magnetic materials). Especially with the current urgent need of automatization for the control of microrobots [3,30], this offers a simple and continuous method to change the swimming behavior drastically by only changing one parameter (frequency) in the system. Path and time optimization algorithms can benefit from micropropellers with frequency-induced reversal of swimming direction, since it facilitates addressing single propellers in a swarm of microswimmers. In this context, it is to keep in mind that with actual practical applications, the requirements on the magnetic fields will increase and set limits to available frequencies and field strengths that could drastically differ from the conditions currently used in labs (e.g. by magnetic coil systems big enough for human limbs). FIRSD- and in general non-linear propellers can therefore be of some advantage for certain challenges. Another challenging task in this regard is to implement a feedback control/visualization of the microswimmers to make automatization possible in the first place. There have been studies using the fields of magnetic resonance imaging (MRI) to power magnetic microswimmers [31,32]. And with magnetic particle imaging (MPI) [33] another visualization method is on the horizon that might facilitate this for medical applications. In those systems, the magnetic fields possibilities are limited in the sense that they are optimized for the imaging rather than the propulsion. It will therefore be necessary to provide the needed complexity and flexibility for navigation in biological environments by introducing different actuation modes like reversing the swimming direction and changing speed rather by the design of the magnetic micropropellers than by the limited accessible magnetic fields.

For expanding the range of possible new applications, it is crucial to acquire a more substantial knowledge about arbitrary shaped propellers to be able to systematically design micropropellers for specific tasks. Additionally to the two examples shown here, more propellers with frequency-induced reversal of propeller direction were found during the measurements (SM movies). While it might not be a surprise that random shaped micropropellers show deviations from the linear frequency-velocity relationship, those are often very small and negligible [34]. With this, we can already formulate some basic requirements for FIRSD (cf. SM): (i) the magnetic moment of such propellers has to be non-parallel and non-perpendicular to the *principle axis of rotation* [16]; (ii) the coupling matrix has to have elements of opposite sign for at least directions which are determined by the frequency dependend axis of rotations. However, looking at more non-linear cases and their mobility matrices might enable us to draw a clearer map, which geometric features give contributions to such coupling and therefore to the reversal in swimming direction.

Recent progress in material synthesis, especially in 3D printing, already provides tools to potentially realize the production of experimentally found propeller geometries and therefore access to those new actuation strategies, without relying on filtering more randomly shaped propellers. However, not every detail of the observed behavior is currently fully understood. In particular, the Euler angles, which describe the frequency-dependent orientation of the propellers, become time dependent after $f_{so}$ and even though some suggestions have been proposed [14] to describe the interactions in this regime, it cannot explain the here measured experimental data. The option for future applications to be able to design propellers with distinct geometry that actively makes use of the changing axis of rotation or even the observed branching, by actively determining it, can be beneficial. They could either swim in opposing direction, use the non-linear regime to increase or decrease the effective coupling and therefore speed up or slow down the velocity, or simply overcome certain setup limitations in regard of the magnetic field. This will therefore help to facilitate and speed up envisioned tasks and is a step towards user and application-friendly micropropellers.

# Apendix

## A. Estimation of rotational drag coefficients and magnetic moment

The magnitude of the magnetic moment and its angle towards the long axis of the chosen propeller coordinate system can be estimated with the experimentally measured characteristic frequencies $f_{tw}$ and $f_{so}$ (or the according angular velocities $\omega_{tw}$ and $\omega_{so}$) from the relevant drag or friction coefficients [13,14]. Therefore an approximation of the propellers of either a cylinder [13,21] or an ellipsoid [14] is needed, to analytically calculate the rotational drag coefficients. We chose to follow Ortega and de la Torre [21] and Ghosh *et al*. [13] since the cylindrical approximation seemed rather fitting for our propeller shapes (cf. microscope images and the 3D-reconstruction of the exemplary propellers in Fig. 1). The according drag coefficients for rotations around the short ($\gamma_s$) and long ($\gamma_l$) cylinder axis are as follows:

$$\gamma_s = \frac{\pi \eta l^3}{3 (\ln p + C_r^\perp)} \qquad (1)$$

$$\gamma_l = \frac{\pi \eta l^3 (1 + C_r^\parallel)}{0.96\, p^2} \qquad (2)$$

with

$$C_r^\perp = -0.662 + \frac{0.917}{p} - \frac{0.100}{p^2} \qquad (3)$$

$$C_r^\parallel = \frac{0.677}{p} - \frac{0.183}{p^2} \qquad (4)$$

where $\eta = 8.9 \cdot 10^{-4}$ Pa s is the viscosity of water at 25°C and $p = l/d$ the ratio of the cylinder length $l$ and its diameter $d$. Then the component of the magnetic moment fixed in the propeller along the long axis of the cylindrical approximation is

$$m_\parallel = \frac{\omega_{tw}}{B_0 \gamma_s} \qquad (5)$$

and along the short axis

$$m_\perp = \gamma_l/B_0\sqrt{\omega_{so}^2 - \omega_{tw}^2}. \tag{6}$$

The magnitude of the magnetization and its angle towards the long axis then is:

$$m = \sqrt{m_\parallel^2 + m_\perp^2} \tag{7}$$

$$\Phi = \arctan m_\perp/m_\parallel \tag{8}$$

Details on the reference frame of the cylindrical approximations and the azimuthal angle can be found in the supplemental material (Figure S5). The values for the magnitude of the calculated magnetic moments of propeller 1 and 2 based on this approximation are shown in Table S1, together with the quantities required for the calculation (magnetic field strength $B_0$, characteristic propeller length $l$ and diameter $d$, characteristic frequencies $f_{tw}$ and $f_{so}$). The magnetic moment can additionally be compared with the magnetization saturation of the respective propeller assuming the same volume of ordered material (Table S1). To calculate the relative magnetization of the propellers the saturation magnetization of maghemite [22] is needed ($M_s = 380$ kA m$^{-1}$):

$$m_{sat} = \pi L \frac{d^2}{4} M_s. \tag{9}$$

The relatively low magnetization values of 0.3 % and 2 % for propeller 1 and propeller 2 respectively can be explained since the random shaped propellers consist of unordered maghemite nanoparticles.

**B. Theoretical framework for arbitrarily shaped magnetic micropropellers**

Below the step-out frequency $f_{so}$, it is possible to describe the orientation of a propeller by the Euler angles $\phi$, $\theta$ and $\psi$ [35]. Their dependency on the frequency is determined by the propeller shape and the orientation of the magnetic moment with respect to this shape. The propeller geometry determines the mobility matrices: the rotational mobility matrix **F**, the translational mobility matrix **ε** and the coupling mobility matrix **G** [16]. A so called center of hydrodynamic mobility can be found for every propeller geometry, where **F** is diagonal and **G** is symmetric, [16] similar to the center of hydrodynamic reaction described by Happel and Brenner [36]. Together with the magnetic moment (magnitude, orientation) and the magnetic field strength, **F** determines the Euler angles and therefore the orientation of the propeller, depending on the applied frequency (cf. next paragraph on mobility matrices and lit. [16]), whereas **G** couples the forced rotation to an effective translation of the propeller. According to Morozov *et al*, [16] the resulting velocity $v = v^{(I)} + v^{(II)}$ along the axis of rotation can be written as the sum of the velocities coming from the diagonal and off-diagonal elements of **G**, $v^I$ and $v^{II}$ respectively:

$$\frac{v^{(I)}}{\omega l} = Ch_1 \sin^2 \psi \sin^2 \theta + Ch_2 \cos^2 \psi \sin^2 \theta + Ch_3 \cos^2 \theta \qquad (10)$$

$$\frac{v^{(II)}}{\omega l} = Ch_{12} \sin 2\psi \sin^2 \theta + Ch_{13} \sin \psi \sin \theta + Ch_{23} \cos \psi \sin 2\theta \qquad (11)$$

with the length $l$ of the propeller, the frequency of the external field $\omega = 2\pi f$ and $Ch_i = G_{ii}/F_i$ and $Ch_{ij} = \frac{1}{2l}(G_{ij}/F_j + G_{ji}/F_i)$ when $i \neq j$.

The advantage of the cylindrical approximation is that it also simplifies the calculation of the frequency-dependent Euler angles (see next paragraph). The transverse rotational isotropy of the propellers [16] basically implies a similar magnitude of the rotational mobility coefficients in the non-elongated directions, as it is given for a approximated cylinder: $F_1 = F_2 < F_3$. Therefore, using this approximation leaves only the elements of the mobility coupling matrix $G_{ij}$ as unknown parameters (cf. Eq.(10-11)), which are used as fitting parameters on the experimental data. However, the more ideal way would be to have access to the geometrical parameters *G* and *F* (e.g. through simulation of the reconstructed shape) and only use the magnetic moment and the characteristic frequencies as fit parameters, which is work in progress.

**Mobility matrices and Euler angles**

The mobility matrices, determined by the geometry of the propeller, relate forces ($\mathcal{F}$) and torques (**L**) with velocities (**U**) and roation rates ($\mathbf{\Omega}$) for the low Reynolds number:

$$\begin{pmatrix} U \\ \Omega \end{pmatrix} = \begin{pmatrix} \varepsilon & G \\ G^\dagger & F \end{pmatrix} \begin{pmatrix} \mathcal{F} \\ L \end{pmatrix} \quad (12)$$

They have the following form in the center of hydrodynamic mobility [16], which axes were chosen here to be along the long and short axis of the approximated cylinder:

$$\begin{pmatrix} F_1 & 0 & 0 \\ 0 & F_2 & 0 \\ 0 & 0 & F_3 \end{pmatrix} \quad (13)$$

$$\begin{pmatrix} G_{11} & G_{12} & G_{13} \\ G_{12} & G_{22} & G_{23} \\ G_{13} & G_{23} & G_{33} \end{pmatrix} \quad (14)$$

With the Euler angles $\phi$, $\theta$ and $\psi$ describing the orientation of the propeller in the lab system (cf. Fig. S3.) [35] and the two angles $\alpha$ and $\Phi$ that describe the orientation of the magnetic moment in the propeller, the transverse rotational isotropy solution is accessible for the cylindrical approximation. $F_\perp = F_1 = F_2 < F_3 = F_\parallel$ then applies for the three diagonal elements of **F**, with $F_1$ and $F_2$ being the inverse rotational drag coefficients ($\gamma_s^{-1}$) of the cylinder rotating around the short axis and $F_3$ the inverse rotational drag coefficient around the long axis ($\gamma_l^{-1}$), respectively. In this case, the solution for the frequency dependent Euler angles can be given for the two synchronous regimes: tumbling solution before $\omega_{tw}$ [16]:

$$\theta = \frac{\pi}{2}, \qquad \psi = -\alpha, \qquad \tilde{\phi} = -\Phi + \arccos \tilde{\omega} \quad (15)$$

wobbling solution between $\omega_{tw}$ and $\omega_{so}$ [16]:

$$\theta_1 = \arcsin \frac{\cos \Phi}{\tilde{\omega}}, \psi_1 = -\alpha - \arcsin \frac{\cos \theta_1 \tilde{\omega} F_\perp}{\sin \Phi F_\parallel}, \quad \tilde{\phi}_1 = 0 \quad (16)$$

$$\theta_2 = \pi - \theta_1, \quad \psi_2 = -2\alpha - \psi_1, \quad \tilde{\phi}_2 = 0 \quad (17)$$

with $\widetilde{\omega} = \frac{\omega}{m B_0 F_\perp}$ and $\phi = \tilde{\phi} - \omega t$.

The second values for the wobbling solution correspond to the second branch solution. Together with Eqn. (10) and (11), the frequency dependent velocity for the transverse rotational isotropic approximations is accessible [16].

## Acknowledgements

The authors thank Prof. Stefan Klumpp for discussion, Dr. Mathieu Bennet and Klaus Bienert for technical advice and support.

**Funding:** This work was funded by Deutsche Forschungsgemeinschaft within the Priority Program 1726 on microswimmers (grant No. FA 835/7-1) and the Max-Planck Gesellschaft. A.C. is supported by the IMPRS on Multiscale Biosystems.

# Supplemental Material

## Using Shape Diversity on the way to new Structure-Function Designs for Magnetic Micropropellers

*Felix Bachmann, Klaas Bente, Agnese Codutti, Damien Faivre\**

### I. Experimental setup and procedure

The used randomly shaped micropropellers consist of iron(III)-nanaoparticles (20-40nm, NanoArc®, Alfa Aesar) that were connected via hydrothermal carbonization to rigid structures with a magnetic moment fixed to their geometry[1]. Stored in a dilute suspension in deionized water, some propellers are filled into a glass capillary (0.2x2x50 mm) by capillary forces: the suspension is sucked into the capillary and the open ends are sealed with petroleum jelly to hinder evaporation and unwanted flows of the fluid. The capillary is fixed with sticky tape on an objective slide, which is, in turn, put on the stage of a custom made Helmholtz-Coil setup[2]. Briefly, this consists of three perpendicular pairs (x, y and z) of parallel coils that are able to produce homogeneous magnetic fields between them (**Figure S1**). An easy way to obtain a rotating magnetic field (in the x-z-plane) is by applying a sinusoidal signal on two of those coils (e.g. x and z), with a phase difference of 90° between them. The resulting used magnetic field in this setup is described by

$$B(\omega, t) = \begin{pmatrix} B_0 \sin(\omega, t) \\ 0 \\ B_0 \cos(\omega, t) \end{pmatrix} \qquad (1)$$

with the angular frequency $\omega = 2\pi f$ and *f* being the frequency of the magnetic field. The setup is able to provide a magnetic field strength up to 4 mT, depending on the frequency of oscillation. The magnetic field strength is typically chosen to be between 1 and 3 mT and the frequency between 1 and 150 Hz depending on the characteristics observed during the measurement. The capillary hangs up side down on the objective slide and is illuminated from the top by a LED-light source (either 400 nm or 635 nm, CoolLED Ltd.). Below the capillary, a 60x Plan Apochromat 1.20 WI Nikon® objective is placed in the optical path, which is led by a mirror system to a high-speed camera, where the images are recorded and saved to a connected computer. The two cameras used for these measurements are an Andor Zyla 5.5 sCMOS and an Optronis CR3000x2 with a maximum resolution of 2560 x 2160 and 1710 x 1696 pixels respectively. Depending on the chosen resolution and binning, up to around 540 frames per second (fps) are possible. Normally the images were recorded with 51 fps. The propellers typically accumulate at the bottom surface of the capillary in the absence of a field due to gravitation, or if no adequate magnetic field is applied. By choosing a magnetic field rotating in the xy-plane, some of the random shape propellers meet the requirements to swim upwards along the z-direction to the middle of the capillary. Here, the rotating field is switched to the xz-plane, which also switches the propelling direction to be along the y-axis. The measurements are performed in this configuration. It is possible to find propellers that reverse their swimming direction depending on the frequency by testing different frequencies (usually in 10 Hz steps). Once a candidate propeller is found, it is moved above the middle (in z-direction) of the capillary as described before where the walls are still far away to not or at least not significantly influence the propeller movement (e.g. rolling over the surface). The propeller is then brought in the approximately (see discussion below) same orientation for every frequency by applying a constant magnetic field along the x-direction before $B(\omega,t)$. The rotating magnetic field the is applied for 10 s for a certain frequency and the propulsion of the propeller is imaged and recorded. After the constant field along x is applied again, a rotating field with a frequency increased by 1 Hz is directly set, so that there is no frequency sweep, but discrete measurements for every frequency. This measurement scheme is repeated to cover the mentioned frequency regime. Since the propellers eventually sink too much towards the lower capillary surface between the measurements (due to gravity), it has to be brought back towards the bulk fluid. After all measurements, a start and end position can determined from the recorded images. The starting point is set 2 s after the start of the rotational field and the end point 2 s before the field was switched off again to ensure that those positions are within the actuation time of 10 s. Therefore, a typically 6 s period of swimming time was recorded. The velocity along the axis of rotation for each frequency is calculated by taking only the distance in y-direction during this time. Close to the step-out frequency, the propellers sometimes started with a rotation around their short axis but after some time (mostly less than 2 s) switched to a rotation around their long axis and remained in this state. This seems to be an effect of the bi-stability in proximity to the step-out frequency[3]. Only the longer of the two possible behaviors was taken into account in order to get a representative velocity average over time. A switch between those configurations was not observed during the measuring time above a certain frequency. This frequency marks a point where the propeller only propelled in an asynchronous way and was therefore taken as the step-out frequency $f_{so}$. Similarly, the transition frequency $f_{tw}$ is extracted from the orientation of the recorded images and the measured data points: the end of the linear regime and the start of the propeller tilting towards the axis of rotation marks the transition from tumbling to wobbling and therefore from the linear to the non-linear regime. While the measured data points support the estimation of $f_{tw}$ for propeller 2 (clear end of the linear regime), one can argue about a different choice for propeller 1, where the transition from a rather linear regime to the wobbling regime is not as clear. Therefore the first frequency, where a notable precession angle change was obvious, is set as the transition frequency.

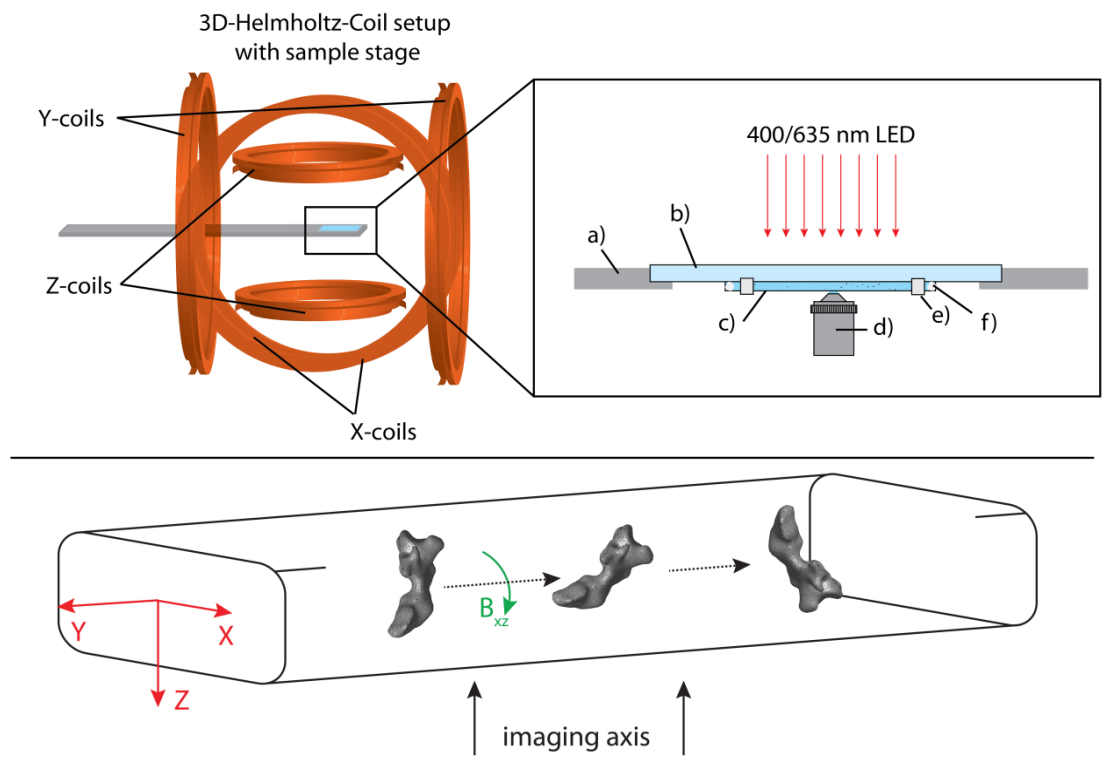

**Figure S1. Schematic experimental setup.** Three perpendicular Helmholtz-coils are generating a homogeneous field around a sample fixed on the sample stage (a). The capillary filled with propeller suspension (c) is fixed with sticky tape (e) on an objective slide (b). To prevent evaporation, the capillary is sealed with petroleum jelly (f). The sample is illuminated from above by either 400 nm/blue or 635 nm/red LED-light and imaged from below by a Plan Apo 60x 1.20 WI objective. The glass capillary (0.2 x 0.2 x50 mm) is filled with di-ionized water and several µm-sized propellers are actuated by a magnetic field rotating in the xz-plane and therefore propel far away from walls along the y-axis (axis of rotation).

**II. Implications of cylindrical approximation and bifurcation**

The curves in Figure 1 A and C show that the measured data can be fitted with the used theory (the branching is included in the fit of the measured data) and reflect the branching behavior between $f_{tw}$ and $f_{so}$ as well. For propeller 1, there are deviations from the theoretical curve. For propeller 2, the theory qualitatively fits better to the experimental data. This can be due to an improved applicability of the approximations and assumptions made (cylindrical shape). The transverse rotational isotropy assumption was adequate to describe the two here reported propellers and the cylindrical approximations allowed for a quick estimation of the magnetic moment (magnitude and angles). But for other propellers, this might not be sufficient and the full solution with mobility matrices derived from simulations of the propeller shapes might be better suited to describe the dynamics in an applied rotating magnetic field[4]. In the rather crude approximation used, the key elements of the propeller behavior are nevertheless captured (linear tumbling regime, bifurcation in wobbling regime). While many propeller designs might aim for a stable propulsion, insusceptible for perturbations and therefore want to prevent bifurcation (e.g. by a certain orientation of the magnetic moment[4]). This might also offer new possibilities: for certain starting conditions (e.g. through a constant field that puts the propeller in a similar starting configuration as it has during rotation), it might become possible to determine the solution branches of the propeller and therefore to have the option of two different swimming velocity responses for one frequency. For some cases, this can be used as an additional way to swim in two opposing directions at the same rotation direction of the external field or in other cases to simply change between two different velocities (e.g. slow and fast mode) for a constant field frequency. Close to the step-out frequency $f_{so}$, even three different modes and velocities might be possible (cf. propeller 1 at ≈ 55 Hz, Fig. 2 A).

When measuring the velocity-frequency relationship of the same propeller multiple times, it happened that for a certain frequency the measured points were on different braches. In **Figure S2** four runs of propeller 2 are shown, where no clear pattern of the population of the branches can be seen. Although the number of experiments is rather small, this indicates the influence of small perturbations (like thermal noise) on the distribution.

This can be supported by looking at the starting positions of the propeller for the respective data points. For propeller 2 an example is shown in **Figure S3**. Here, the propeller starting condition before the rotating field is applied is shown. Although the same constant field (vertically, red arrow) was applied on both images (left, just before 87 Hz; right, just before 90 Hz), the propeller still has some degree of rotational freedom and can therefore e.g. through thermal noise and diffusion obtain both configurations shown in the images. These two different starting conditions led to two different velocity responses/branches in the afterwards applied rotating field. However, not all cases/starting positions were as clear as in the two images shown, but still, this shows that thermal noise probably has a strong influence in the determination of the branch. Since the rotational diffusion scales inversely with the rotational drag coefficient, it could be that the bigger diameter of propeller 1 and therefore the bigger drag coefficient and smaller diffusion coefficient lead to smaller fluctuations of the starting position in the given time frame for propeller 1 compared to propeller 2. As a consequence, the population of the two branches for propeller 2 is more equally spread compared to propeller 1. A potential solution could be applying a constant field in horizontal direction (which means not directly in the middle between the solution configurations as above). Then this would already be a similar configuration as in one of the two solution branches and could therefore limit the possibility of the propeller to take the solution branch with the magnetic moment anti-parallel to before applied constant field, when the rotating magnetic field is applied.

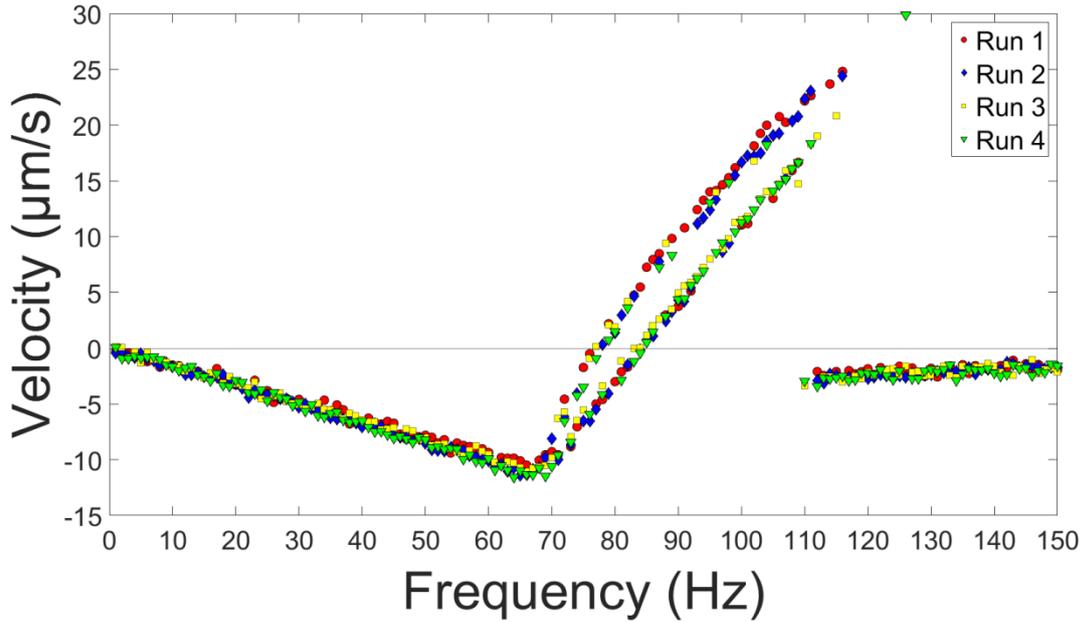

**Figure S2. Branching.** Propeller 2 was measured four times for frequencies between 1 and 150 Hz. While the general behavior was nearly identical for each run, the population of the two branches between $f_{tw}$ and $f_{so}$ varied in those measurements.

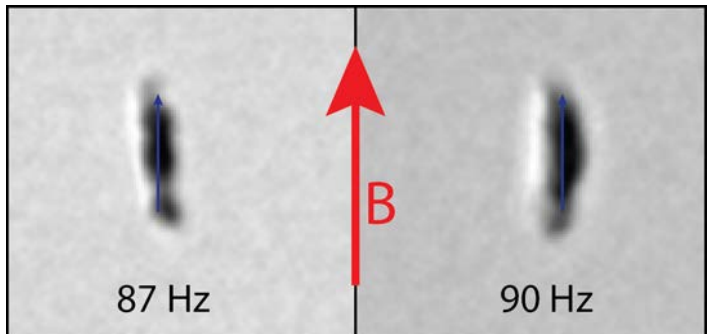

**Figure S3. Propeller starting positions.** Propeller 2 at two different frequencies (87 Hz, left and 90 Hz, right). The magnetic moment (blue) is roughly along the long axis of the propeller. If a constant magnetic field (red) is applied, the propeller can still rotate around the axis of the magnetic field. This enables two different starting conditions that led to two different velocity responses (branches) during the measurement.

### III. Euler angles and coordinate systems

Following Morozov et al.[4], in the synchronous regime, where the balance of hydrodynamic and magnetic torque is still valid, the orientation of a body-fixed coordinate system (BCS) in the laboratory coordinate system (LCS) can be described by using the frequency dependent

Euler angles $\phi$, $\theta$ and $\psi$ (cf. **Figure S4**.)[5]. This dependency is determined by the mobility matrix **F**. The mobility matrix determines the hydrodynamic torque acting on the propeller. The characteristic frequencies (when a propeller starts to wobble and when the asynchronous regime is reached) are determined by *F* and the magnetic moment. We have access to the characteristic frequencies from the measurements and to the drag coefficients of the propeller through the cylindrical approximation. This in turn allows the estimation of the magnetic moment in the propeller (magnitude and angles $\alpha$ and $\Phi$). Here, we take the knowledge of the low-frequency solution[6] into account: in this regime, the magnetic moment rotates in plane with the magnetic field. Our reference coordinate system is therefore an image of the respective propeller at low frequencies (**Figure S5**), where the 2D-projection of the propeller is largest. This projection is approximated with a cylinder of length l and diameter d (white box) and the body coordinate frame axes are along the cylinder axis (in red, $x_1$, $x_2$ and $x_3$). In this coordinate frame, the $\alpha$- angle is 0, since the magnetic moment *m* (blue) is in the $x_1 x_3$ - plane and the remaining $\Phi$-angle can be estimated from the characteristic frequencies (cf. next section or lit. [4]).

After estimating the rotational mobility coefficients, assuming $\alpha = 0°$ and estimating $\Phi$, the velocity remains a function of the six elements of **G**. The entries of *G* along the current axis of rotation could be derived from the fit on the measured data by using the Mathematica function NonlinearModelFit.[7] Since for propeller 1, there were only a few points on the second branch, those points (at 34, 36, 37, 39, 40 and 52 Hz) got weighted four times to balance the number of points in the first branch (using the weights-option of NonlinearModelFit) to take both possible solutions equally into account. Points that were already in the after step-out behavior (can be seen on microscope images) were excluded (propeller 1; at 51 Hz; propeller 2: at 112, 113, 114 and 115 Hz). In Tab. S1, all parameters and values for and from the approximations can be found for the two example propellers. Currently, the here observed behavior after $f_{so}$ is not covered by theory: the propellers seem to fall back to a rotation around their short axis (some kind of tumbling). For our examples, there was no intermediate state similar to the wobbling regime, which was also reported for similar cases.[3]

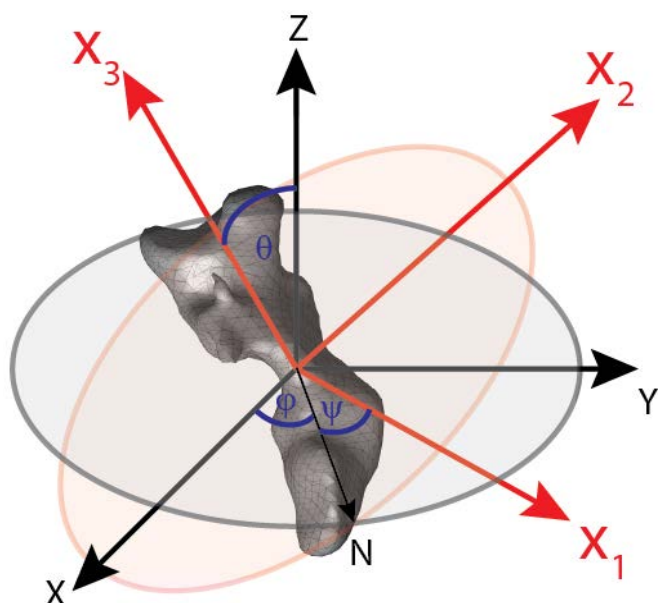

**Figure S4. Coordinate systems.** Schematic of the lab coordinate system (LCS, black) and the body centered coordinate system (BCS, red) connected via the Euler angles $\phi$, $\theta$ and $\psi$.[5]

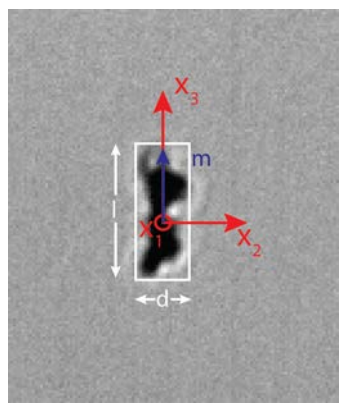

**Figure S5. Cylindrical approximation.** The propeller was approximated by a cylinder (white box) of length l and diameter d. The cylinder axes form a body coordinate system (BCS) with $x_1$, $x_2$ and $x_3$ where the magnetic moment $m$ (blue) is in the $x_1 x_3$ plane.

**Table S1. Measured and calculated properties.**

| Properties | Propeller 1 | Propeller 2 |
|---|---|---|
| $B_0$ (mT) | 2 | 1 |
| $l$ (μm) | 4.34 | 4.13 |
| $d$ (μm) | 1.59 | 1.28 |
| $\gamma_s$ (m² kg/s) | $1.14 \cdot 10^{-19}$ | $8.36 \cdot 10^{-20}$ |
| $\gamma_l$ (m² kg/s) | $3.92 \cdot 10^{-20}$ | $2.36 \cdot 10^{-20}$ |
| $f_{tw}$ (Hz) | 25 | 68 |
| $f_{so}$ (Hz) | 55 | 117 |
| $m$ (Am²) | $1.08 \cdot 10^{-14}$ | $3.84 \cdot 10^{-14}$ |
| $\Phi$ (°) | 34.11 | 21.60 |
| $\alpha$ (°) | 0 | 0 |
| $m/m_{sat}$ | 0.003 | 0.02 |
| $F$ (m$^{-2}$ kg$^{-1}$ s) | $10^{18}\begin{pmatrix} 8.81 & 0 & 0 \\ 0 & 8.81 & 0 \\ 0 & 0 & 25.5 \end{pmatrix}$ | $10^{18}\begin{pmatrix} 11.97 & 0 & 0 \\ 0 & 11.97 & 0 \\ 0 & 0 & 42.32 \end{pmatrix}$ |
| $G$ (m$^{-1}$ kg$^{-1}$ s) | $10^{10}\begin{pmatrix} 8.568 & 52.22 & -3.24 \\ 52.22 & 21.43 & -39.61 \\ -3.24 & -39.61 & 6.45 \end{pmatrix}$ | $10^{10}\begin{pmatrix} 38.81 & -5.23 & -21.61 \\ -5.23 & -32.12 & 5.83 \\ -21.61 & 5.83 & 70.01 \end{pmatrix}$ |

**IV. Reversal of swimming direction and dimensionless velocity**

The experimental data shows that it is possible to reverse the swimming direction of specific micropropellers by changing the applied external magnetic field frequency. Those micropropellers change their axis of rotation with frequency, which in turn leads to different rotation-translation-coupling coefficients with, in this case different signs, resulting in swimming in two opposing directions. The dimensionless velocities $U = 1000 \cdot v/(l \cdot f)$ of the analyzed propellers are similar to state-of-the-art fabricated magnetic microswimmers[8, 9, 10, 11, 12, 13, 14]. Additionally, propeller 1 approaches $U = 100$ for the measured points falling on the second branch (green dashed line in Fig. 1 A). On the one hand, this confirms that selected swimmers from a pool of random shaped micropropellers may be fast swimmers[15]. On the other hand, this shows that effective propulsion with the same propeller in two different directions can be done by only changing the frequency of the applied magnetic field, even without relying on the different branches between $f_{tw}$ and $f_{so}$.

## V. Towards maximizing the utilization of geometry

For every propeller geometry, there is at least one axis of rotation, where the coupling between rotation and translation is best, which means, the dimensionless speed *U* as a measure of this coupling is the highest. This applies for positive and negative values of *U*. It is possible to get the two extreme values of U for each geometry, given that the matrices **F** and **G** are known, by finding the axis of rotation vectors $\widehat{\Omega}^{BCS}$, which maximizes and minimizes the quadratic form of the coupling matrix[4] $U \propto \widehat{\Omega}^{BCS} G F^{-1} \widehat{\Omega}^{BCS}$. Looking at the mobility matrices shown by Morozov et al.[4] for helices with 1, 1.5 and 2 turns, the highest dimensionless velocities are not reached by a rotation around the long axis of the helix but rather with an angle of around 34°, 10° and 31° between the long axis of the helix and the axis of rotation for dimensionless velocities of 407, 248 and 250 respectively. For all three cases, the minimum (negative) dimensionless velocity is found for a rotation around a short helix axis: nearly 90° for 1 or 2 turns and around (80°) for 1.5 turns. The respective dimensionless velocities are -198, -147 and -102. This shows that when just looking at the dimensionless velocities, it seems reasonable to assume helices to be good swimmers in two opposite direction for different frequencies. And indeed Morozov et al. provide frequency-velocity diagrams that show simulations of helices swimming in two opposing directions for special frequencies. The effective propulsion speed (in µm) for negative velocities is rather small and therefore probably not usable for applications. The same treatment can be done for the two propellers presented in this paper. An axis can be found for propeller 1, where the maximum dimensionless speed is -122 while the minimum is -72. For Propeller 2 the maximum dimensionless speed is 59 and the minimum is -41. These values state a limit for the possible dimensionless speed accessible with these respective geometries. The experimentally measured values are within those limits for both propellers (propeller 1: 98 and -52; propeller 2: 51 and -41). It can therefore be helpful to design propellers with the appropriate magnetic moment to exactly rotate around those best coupling axes at a certain frequency. The eventual speed can, to a certain extent, be adjusted via the magnitude of the magnetic moment or the magnetic field: the higher they are, the higher the applied frequencies can be to show the same characteristics and therefore the speed increases. All in all, the non-linear relation and bifurcation of the frequency-velocity response is not just an issue to overcome, but can be actively used to increase the control parameters and possibilities, for magnetically driven propellers.

**VI. Basic requirements for FIRSD**

In general, a propeller needs to meet some basic requirements in order to be able to reverse direction either via changes of the magnetic field frequency or strength. The main requirement is that the propeller actually swims. Thus, it needs a minimum of asymmetry in its geometry that allows coupling between rotation and translation and therefore propulsion[16]. Additionally, this asymmetry leads to a non-linear frequency-velocity response in a rotating magnetic field if the magnetic moment is not perpendicular to the axis of rotation around the hydrodynamically best configuration[17]. In case of an cylindrical object, a magnetic moment perpendicular to the cylinder axis would lead to only one fixed axis of rotation from 0 Hz to the step-out frequency, with a constant coupling between rotation and translation (linear v-f-dependence)[17]. But exactly this change of the rotation axis is needed to achieve frequency-depending rotation-translation coupling (apart from shape deformation[18] and changes to the magnetic moment). The details of this frequency dependent coupling (and therefore the mobility coupling matrix **G**) determine if the propeller is able to reverse its swimming direction for two different frequencies. While these criteria are given for many magnetic microswimmers or can easily be achieved, the absolute and dimensionless velocities as well as their directions remain a function of the specific geometry (**G**) and the associated magnetic moment[4].

**VII. Comparison between linear and non-linear magnetic micropropellers**

It is difficult to make statements regarding the simplicity and effectiveness of magnetic micropropellers and their actuation schemes. As mentioned before, already a simple comparison between velocities does not live up to the more complex reality (magnetic moment orientation, environmental conditions, etc.).

However, when going to very basic and simplified concepts of control and actuation, some differences between linear and FIRSD or in general non-linear micropropellers become apparent. The most basic concept is to break down any 3D movement or actuation into sequences of 1D steps, where a propeller moves along a line from point $x_0$ to $x_1$ and to consider a minimalistic swarm as two propellers. For some applications and here, in these theoretical contemplations, it necessary to bring those two propellers from arbitrary start positions (e.g. both start at $x_0 = 0$) to arbitrary end positions ($x_1$ and $x_2$) only by sequences of 1D movements. Due to the 'global' homogeneous magnetic actuation field, the propellers are all exposed to the same driving torques. Therefore, it is necessary, that the propellers are non-identical to reach positions with $x_1 \neq x_2$. For purely linear propellers, however, with either different slopes (**Fig. S6** A) or different stepout frequencies (Fig. S6 B) it is necessary to use the non-linear behavior after the stepout frequency to have access to arbitrary $x_1$ and $x_2$. These cases have theoretically been studied by Vach *et al.*[19] and show that in fact from a common starting point ($x_0 = 0$) the two arbitrary end points ($x_1$, $x_2$) can be reached by applying the right sequence of magnetic fields. This is already possible with two different actuation frequencies $f_1$ and $f_2$ that are applied for the time $t_1$ and $t_2$, respectively. However, depending on the arbitrary end positions, this can require long path ways ($d \gg x_1$ or $x_2$) for the propellers and therefore a huge amount of space (e.g. when the end positions are in opposite directions from the start position). Since space can be a limiting parameter in many applications, the way to go is to have many shorter paths instead of two big paths by constantly changing between the frequencies $f_1$ and $f_2$. The total time (apart from time needed to change the magnetic fields, which is assumed to be 0) stays the same and the propeller move directly from their starting to their end position in the limit of infinite small steps.

However, one interpretation of simplicity for the discussed task could be to require the least amount of steps and therefore 'the least amount of user input'. In this sense, the simplest way for this basic task would be to do it in one step, which in turn would mean that the propellers take the shortest way from start to finish. The ratio of accessible velocities determines the possible end positions ($v_1/v_2 \sim x_1/x_2$). Given the two velocities $v_1$ and $v_2$ at the applied frequencies f for the two considered propellers respectively, arbitrary end points $x_1$ and $x_2$ (this means the quotient of $x_1/x_2$ can take values from $-\infty$ to $+\infty$) can be reached in one step if the quotient of $v_1(f)/v_2(f)$ covers the interval from $-\infty$ to $+\infty$ too. But as can be seen in Fig.S6 A+B, the covered interval of propellers with linear velocity-frequency relation is limited to either positive or negative values and depending on the propellers, the possible outcomes / end positions are even more restricted (even though the non-linear stepout behavior is included). The advantage of FIRSD-propellers now is that in one step, a huge interval of different end positions is covered (guaranteed either left or right unbounded) and for some propellers (e.g. propeller 1), it is even possible to reach arbitrary end positions in a single step Fig.S6 D.

As a consequence the two linear propellers need at least 2 actuation steps to reach arbitrary end positions while the FIRSD propeller of Fig. S6 D would only need one single step. As mentioned in the beginning, complex tasks and applications can be reduced to a sequence of those basic steps and a factor of 2 for each step adds quickly up (and could be of even more importance when considering more than 2 propellers). All of these considerations, however, do not take the time needed for these steps into account. In the currently limited range of tasks that are performed in labs, speed is the decisive factor, making 2 fast steps (linear propellers) instead of 1 slower step (FIRSD-propellers) might be better and therefore these contemplations remain mostly theoretical for now. However, it is to mention, that all of this discussion did not include the stepout behavior of FIRSD-propellers, since too little is known about it for now. The stepout-behavior of linear propellers is what makes arbitrary swarm control and manipulation possible in the first place and the results of our measurements give rise to some hopes that FIRSD-propellers can show not only simpler actuations but additionally faster execution times by using their stepout behavior too. This can be supported by the big velocity gap occurring around the stepout frequency of propeller 2 (cf. main text Fig. 1 B). This gap is an important factor for the time, how fast two propellers can be brought to the arbitrary end positions.

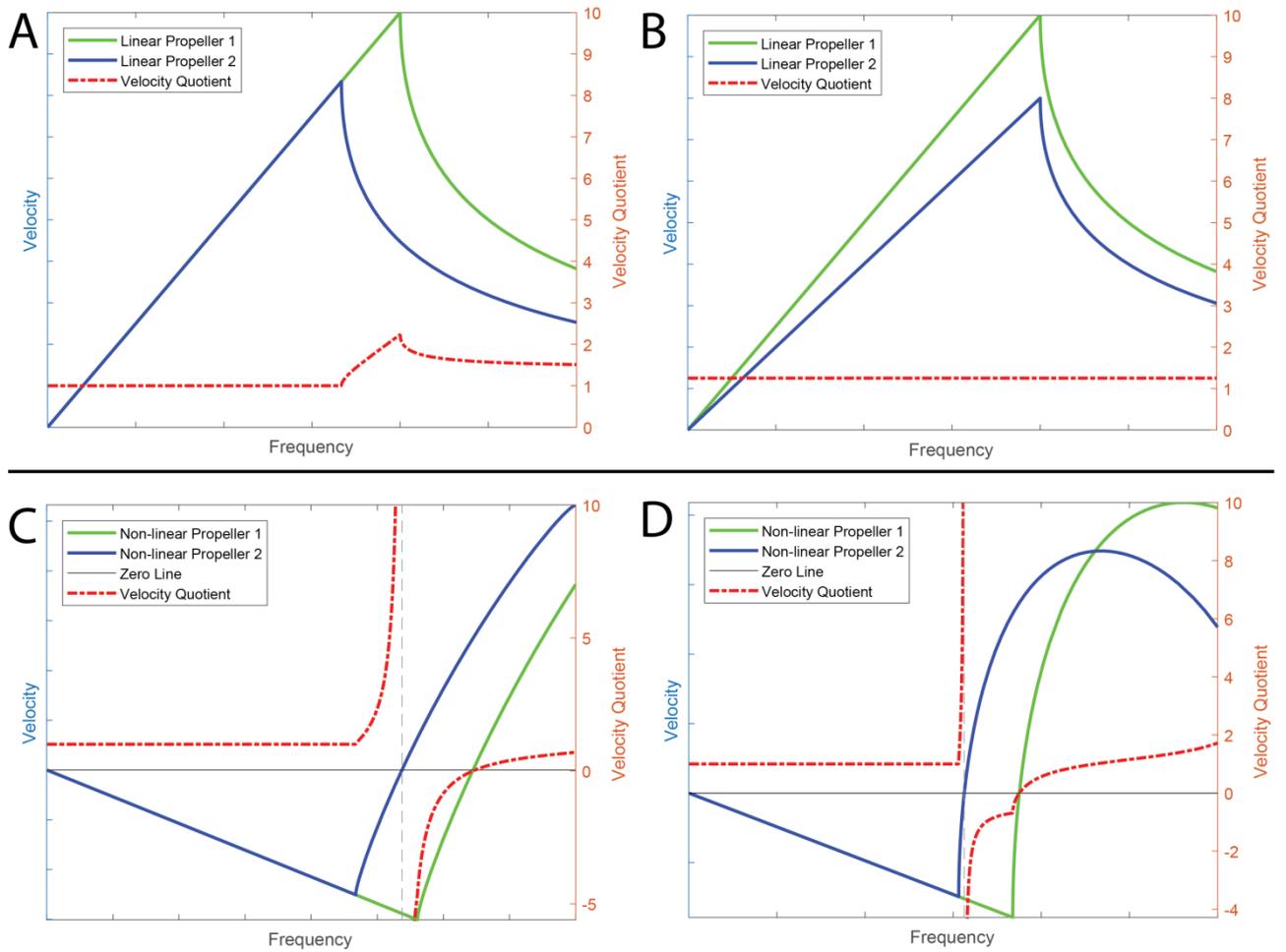

**Figure S6. Linear vs non-linear propellers.** Two typical magnetic micropropellers with a linear v-f-relationship are depicted in A with identical coupling coefficients but different stepout frequencies (e.g. through different magnetic moments, here: $m_1 = 1.2\ m_2$) and in B with identical stepout frequencies but different coupling coefficients. The quotient of both velocity curves (red dotted line) in the respective diagram gives the range of reachable points $x_1$ and $x_2$ in one single actuation step. This range is much smaller than for non-linear and especially FIRSD-propellers. The two propeller behaviors from the main text are shown here. Similarly to the linear propellers, the two curves (blue and green) correspond to different magnetic moments and the red dotted line shows the range of arbitrary points that can be reach from a common starting point by only one actuation step. Already in C nearly all values are covered and the propeller behavior in D shows full cover which means that 2 of those propellers can reach any 2 points from –∞ to +∞. This result is already achieved without regarding the stepout behavior of non-linear propellers that might increase the applicability of FIRSD-propellers.

## VIII. 3D-reconstruction and further examples

There are some methods to acquire 3D information from 2D images. In medicine, x-ray tomography is a common technique to record 1D absorption profiles from different angles to then reconstruct a 3D body with a filtered inverse Radon transformation [20]. While the gray value of 2D bright field microscopy images is not only a function of absorption due to the type and amount of material in the path of the photons but also influenced by the height of the focus plane, reflections and diffractions, it is still possible to get an approximated 3D reconstruction from the 2D gray scale images using a filtered back-projection algorithm. Artificially slicing the 2D images, summing over the different angles and using the matlab-function iradon (filtered back-projection), 3D shapes of the propellers are accessible (more Details: Kak and Slaney [20]). The reconstruction is influenced by the number of images taken (in general, the algorithm is only exact for an infinite amount of images) and the quality of the images as can be seen in **Figure S7**.

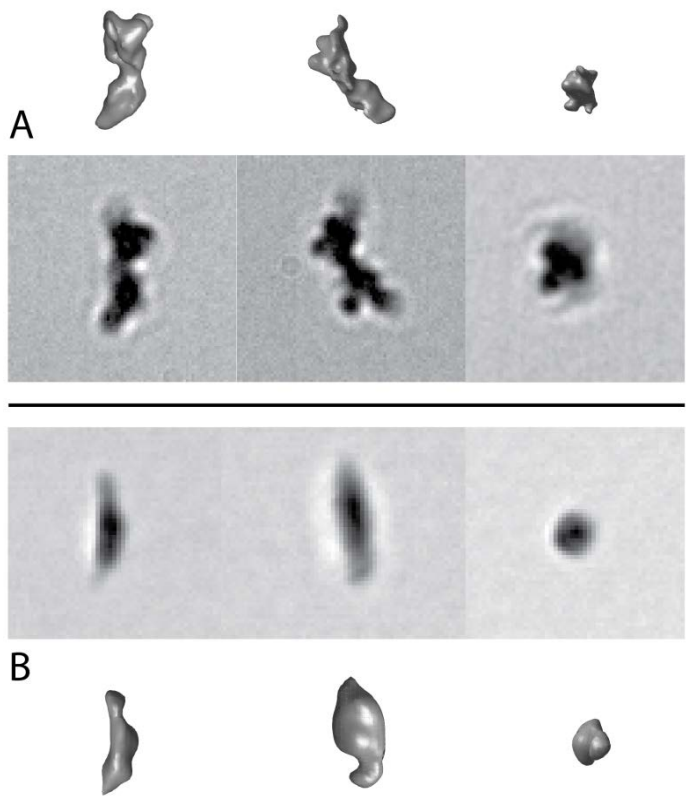

**Figure S7. Propeller projections.** The 3D-reconstructions of propeller 1 (A) and propeller 2 (B) compared with microscope images from similar angles. Although the general shape matches quite well, there are some artifacts arising from limited image number and image quality, especially for propeller 2, where only 51 images were taken into account compared to the 543 images for the propeller 1 reconstruction.

The main focus of the measurements was on receiving the frequency-velocity relationship. Although not all propellers that were found during measurements could be examined in detail as the two examples (propeller 1 (g) and 2 (h)), some more are listed below to give a better impression, on how such propellers could look like. Together with the reconstruction, there are short movies (a)-g)) showing their frequency-induced reversal of swimming direction for two different frequencies with otherwise constant conditions during the respective measurement. The reconstructions vary in quality since not for every propeller images in an adequate number and quality could be recorded, but the reconstructions still give a rough idea about the propellers shapes. It is to notice that most of them are rather elongated, but also some more compact propellers (**Figure S8** c), d) and f)) showed the described behavior. In **Figure S9** some velocity-frequency-measurements for different propellers are additionally shown (Propeller 1 = Example5, Propeller 2 = Example 8). While they also vary in quality (number of frequency measured, influence of thermal noise), they can show the wide range of different behaviors of magnetic micropropellers that have ther magnetic moment not along a principle axis of rotation. Although these examples show a selection bias, they nevertheless show that FIRSD or non-linear v-f-dependencies are rather common and no exception.

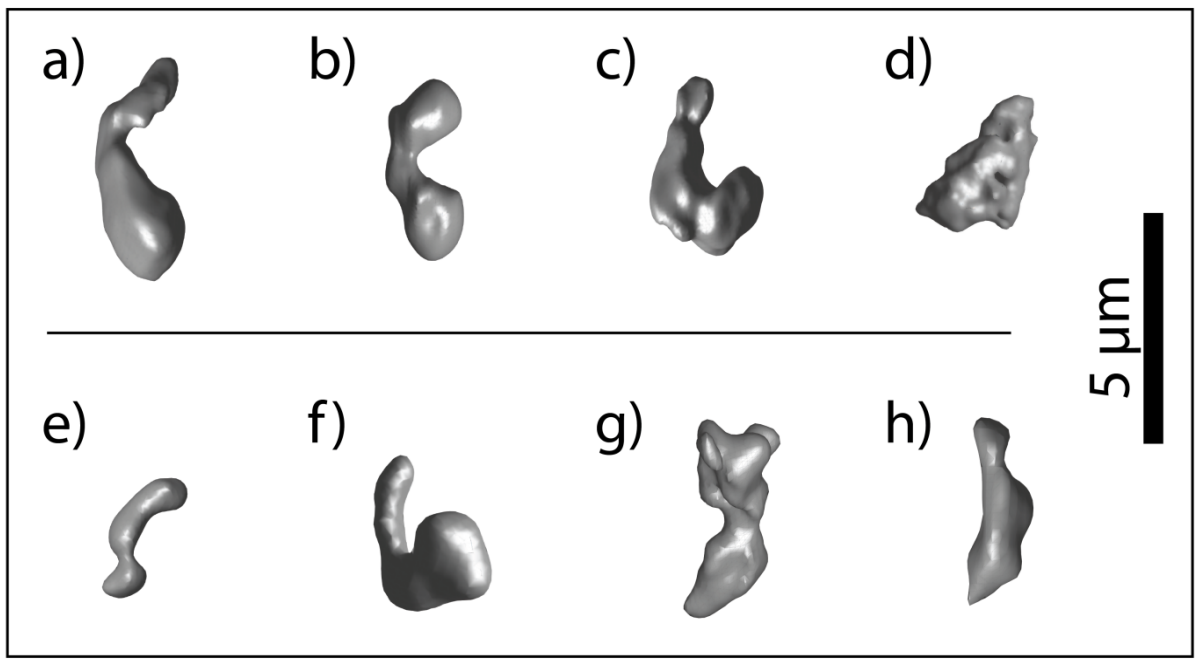

**Figure S8. Propeller examples.** More examples of reconstructed propeller shapes that showed a reversal of propelling direction for certain frequencies of the applied magnetic field.

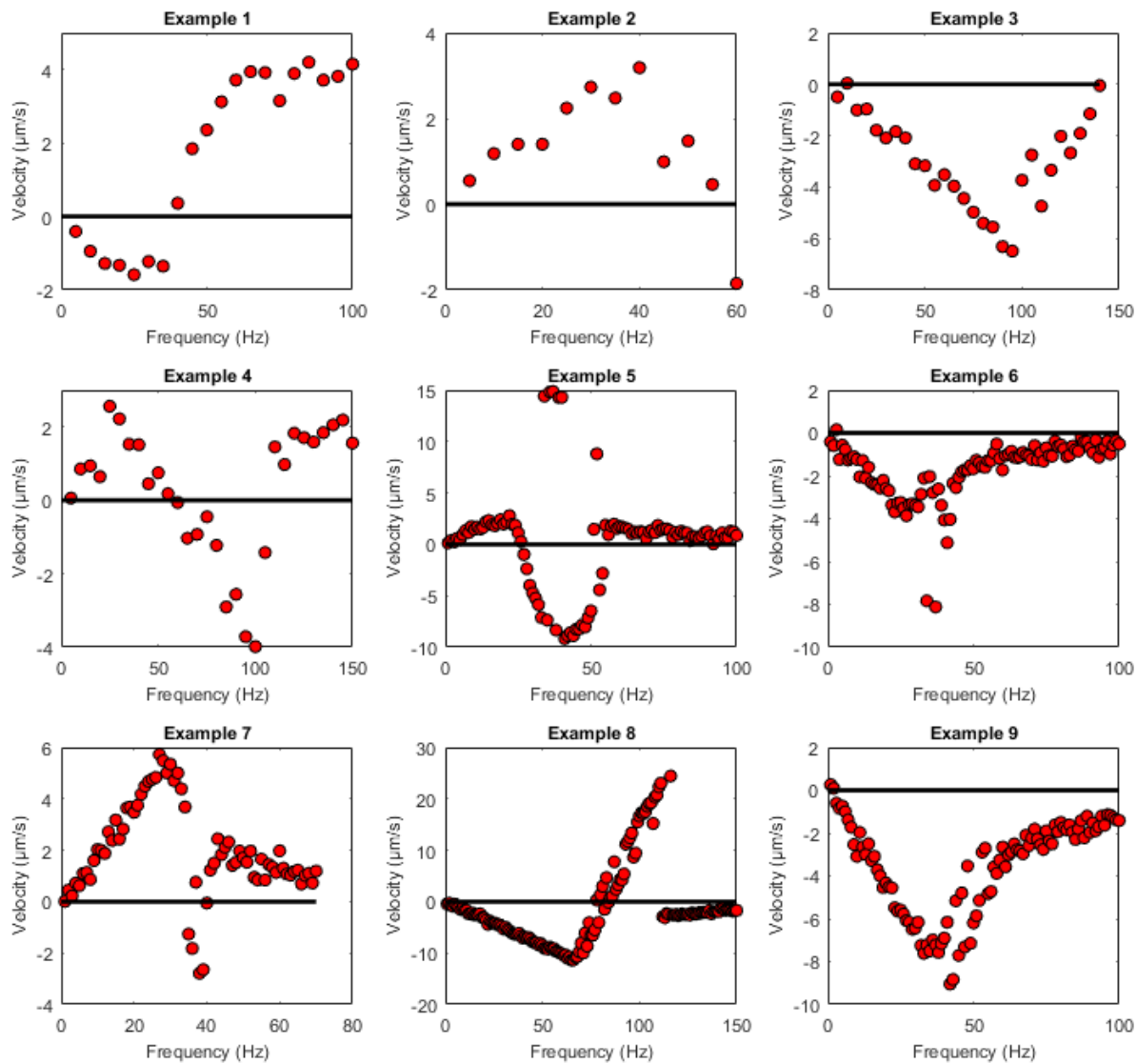

**Figure S9. Propeller Curve examples.** More velocity-frequency-curves have been measured for different propellers. Similar to the reconstructions, they differ in quality but show the quite different behaviors non-linear propellers can show, while on the same time offering comparable features (e.g. three regions: tumbling, wobbling, step out).

**Propeller movie details**

The attached movie files show examples of frequency-induced reversal of swimming direction of propeller a) to h). However, they were recorded under different conditions (frames per second, distance to capillary surface, density of the propeller suspension, applied magnetic field strength and frequencies, image quality, focus plane). Nevertheless, they all show, to different extents, the described behavior (note: the rotation of the propellers is not always clear to see in the following movies due to the used ratio of frequency to frames per second and additionally the movie frame rate):

a) The "boomerang-shaped" propeller rotates around its short axis for $f = 40$ Hz and swims to the right. For 60 Hz it changes its axis of rotation and swims to the left (2 cycles).

b) This propeller has a similar shape and behavior to a): swims to the right for $f = 20$ Hz and to the left for 70 Hz (three cycles).

c) The behavior of this more compact structure seems to be determined by its tail: while for 20 Hz it rotates in the plane of the magnetic field and the propeller swims to the right, the tail behaves more like an actual propeller for 70 Hz and propels the structure to the left for 70 Hz.

d) Even though this propeller is not very slender, it still shows the frequency-dependent propelling direction reversal for 20 (to the right) and 70 (to the left) Hz.

e) Similar to propeller a) and b).

f) Similar structure to propeller c): due to its very compact structure and since this propeller is very small, thermal noise has a strong influence on its behavior. Nevertheless, two clearly opposite directions could be observed, depending on the frequency.

g) This movie shows parts of the measurement of propeller 1 shown in this paper: for 20 Hz it rotates around its short axis and swims rather slowly to the left. For 38 Hz, it rotates around another axis and swims to the right. However, for 39 Hz it swims very fast to the left again despite not rotating around the short axis: the propeller is in the second branch.

h) Here, propeller 2 is shown for 20 Hz and for 101 Hz. The rather fast two opposing swimming directions are observable.

Additionally, with the parameters gained from fitting the measured data, the theoretic behavior of propeller 1 (movie i) and propeller 2 (movie j) is animated, to show the different propeller configuration resulting from theory for different frequencies and branches. Finally, an proof-of-concept it shown in form of movie k). It shows an example of isolating a single propeller from a swarm: for 20 Hz all propellers have a joint movement to the left; at 40 Hz all except one propeller continue to swim to the left, while this one propeller swims to the right allowing a fast isolation and separation and therefore essentially splitting of the swarm.